# Self-assembled tunable photonic hyper-crystals

Vera N. Smolyaninova [1)], Bradley Yost [1)], David Lahneman [1)], Evgenii E. Narimanov [2)], and Igor I. Smolyaninov [3)]

[1)] *Department of Physics Astronomy and Geosciences, Towson University,*

*8000 York Rd., Towson, MD 21252, USA*

[2)] *Birck Nanotechnology Centre and School of Electrical and Computer Engineering, Purdue University, West Lafayette, IN 47907, USA*

[3)] *Department of Electrical and Computer Engineering, University of Maryland, College Park, MD 20742, USA*

**We demonstrate a novel artificial optical material, a "photonic hyper-crystal", which combines the most interesting features of hyperbolic metamaterials and photonic crystals. Similar to hyperbolic metamaterials, photonic hyper-crystals exhibit broadband divergence in their photonic density of states due to the lack of usual diffraction limit on the photon wave vector. On the other hand, similar to photonic crystals, hyperbolic dispersion law of extraordinary photons is modulated by forbidden gaps near the boundaries of photonic Brillouin zones. Three dimensional self-assembly of photonic hyper-crystals has been achieved by application of external magnetic field to a cobalt nanoparticle-based ferrofluid. Unique spectral properties of photonic hyper-crystals lead to extreme sensitivity of the material to monolayer coatings of cobalt nanoparticles, which should find numerous applications in biological and chemical sensing.**

Over the last few decades a considerable progress has been made in developing artificial optical materials with novel and often counterintuitive properties. Revolutionary research by Yablonovitch and John on photonic crystals [1,2] was followed by the development of electromagnetic metamaterial paradigm by Pendry [3]. Even though considerable difficulties still exist in fabrication of three-dimensional (3D) photonic crystals and metamaterials, both fields exhibit considerable experimental progress [4,5]. On the other hand, on the theoretical side these fields are believed to be complementary but mutually exclusive. Photonic crystal effects typically occur in artificial optical media which are periodically structured on the scale of free space light wavelength $\lambda$, while electromagnetic metamaterials are required to be structured (not necessarily in a periodic fashion) on the scale, which is much smaller than the free space wavelength of light. For example, in metal nanowire-based hyperbolic metamaterials [6] schematically shown in Fig.1A the inter-wire distance must be much smaller than $\lambda$. Here we report experimental realization of 3D "photonic hyper-crystals" [7] which bridge this divide by combining the most interesting properties of hyperbolic metamaterials and photonic crystals.

Our concept of the photonic hyper-crystal [7] is based on the fact that dispersion law of extraordinary photons in hyperbolic metamaterials [8]

$$\frac{k_z^2}{\varepsilon_{xy}} + \frac{k_x^2 + k_y^2}{\varepsilon_z} = \frac{\omega^2}{c^2} \tag{1}$$

does not exhibit the usual diffraction limit. In such uniaxial metamaterials the in-plane $\varepsilon_{xy}$ and out-of-plane $\varepsilon_z$ components of the dielectric permittivity tensor have opposite signs (e.g. the metal wire array hyperbolic metamaterial shown in Fig.1A may have $\varepsilon_z < 0$ and $\varepsilon_{xy} > 0$ [8]), so that the photon wave vector components $k_i$ are not bounded at a given frequency of light ω. Existence of large *k* vector modes in a broad range of frequencies is responsible for such unusual effects as hyperlens-based super-resolution imaging [8-11] and broadband divergence of photonic density of states in hyperbolic metamaterials [12]. On the other hand, this also means that periodic modulation of hyperbolic metamaterial properties on a scale $L \ll \lambda$ (see Fig.1B) would lead to Bragg scattering of extraordinary photons and formation of photonic band structure no matter how small *L* is [7]. Thus, so formed "photonic hyper-crystals" would combine the most interesting properties of hyperbolic metamaterials and photonic crystals. For example, similar to classic photonic crystal effect predicted by John [2], strong localization of photons may occur in photonic hyper-crystals. However, unlike usual photonic crystals where light localization occurs on a scale $\sim\lambda$, photonic hyper-crystals may exhibit light localization on deep subwavelength scale. Similar to surface plasmon resonance (SPR) [13] and surface enhanced Raman (SERS) [14] based sensing, engineered localization of light on deep subwavelength scale in photonic hyper-crystals should find numerous applications in biological and chemical sensing.

Band structure and field distribution inside a photonic hyper-crystal may be obtained in a straightforward manner. While both $\varepsilon_{xy}$ and $\varepsilon_z$ may exhibit periodic spatial dependencies, let us consider the relatively simple case of coordinate-independent $\varepsilon_{xy} >$ and periodic $\varepsilon_z(z) < 0$ with a period $L \ll \lambda$. Aside from the relative mathematical simplicity of this model, it also corresponds to the most readily available low-loss realizations of a hyperbolic metamaterials such as the composites formed by metallic nanowires in a dielectric membrane [15] (where $\varepsilon_{xy} > 0$ and $\varepsilon_z(z) < 0$), and planar layered metal-dielectric [16,17] and doped semiconductor metamaterials [18]. Taking into account the translational symmetry of the system in $x$ and $y$ directions, we can introduce the in-plane wave vector ($k_x$, $k_y$) so that the propagating waves can be expressed as

$$E_\omega(\vec{r}) = E(z)\exp(ik_x x + ik_y y) \tag{2}$$

$$D_\omega(\vec{r}) = D(z)\exp(ik_x x + ik_y y)$$

$$B_\omega(\vec{r}) = B(z)\exp(ik_x x + ik_y y)$$

The uniaxial symmetry of this medium reduces the ordinary and extraordinary waves to respectively the TE ($\vec{E} \perp \hat{z}$) and TM ($\vec{B} \perp \hat{z}$) - polarized modes. Introducing the wavefunction $\psi(\vec{r})$ as the $z$-component of the electric displacement field of the TM wave

$$\psi(\vec{r}) = D_z(\vec{r}) = \varepsilon_z(z)E_z(\vec{r}) = -\frac{c}{\omega}k_x B \tag{3}$$

for the wave equation we obtain

$$-\frac{\partial^2\psi}{\partial z^2} + \frac{\varepsilon_{xy}}{\varepsilon_z(z)}\psi = \varepsilon_{xy}\frac{\omega^2}{c^2}\psi \tag{4}$$

In this wave equation the periodic $\varepsilon_{xy}/\varepsilon_z$ ratio acts as a periodic effective potential. As usual, solutions of eq.(4) may be found as Bloch waves

$$\psi(z) = \sum_{m=0}^{\infty}\psi_m \exp(i(k_z + \frac{2\pi}{L}m)z) \tag{5}$$

where $k_z$ is defined within the first Brillouin zone $-\pi/L < k_z < \pi/L$. Strong Bragg scattering is observed near the Brillouin zone boundaries at $k_z \sim \pi/L >> \pi/\lambda$, leading to the formation of photonic bandgaps in both the wavenumber and the frequency domains. This behavior is illustrated in Fig. 2, where we compare the dispersion diagram for an example of such photonic crystal to its effective medium counterpart.

Similar to usual photonic crystals [2], adiabatic chirping of $L$ leads to strong field enhancement which, unlike in usual photonic crystals, occurs on a deep subwavelength scale. We should also mention that the most interesting case appears to be epsilon near zero (ENZ) situation where $\varepsilon_z$ approaches zero near a periodic set of planes. As has been demonstrated in Ref. [19], electric field of the extraordinary wave diverges in these regions. These periodic field divergences appear to be most beneficial for sensing applications.

In order to validate the photonic hyper-crystal concept we have developed an experimental technique which uses three dimensional self-assembly of cobalt nanoparticles in the presence of external magnetic field. Magnetic nanoparticles in a ferrofluid are known to form nanocolumns aligned along the magnetic field [20]. Moreover, depending on the magnitude of magnetic field, nanoparticle concentration and solvent used, phase separation into nanoparticle rich and nanoparticle poor phases may occur in many ferrofluids [21]. This phase separation occurs on a 0.1-1 micrometer scale. Therefore, it can be used to fabricate a self-assembled photonic hypercrystal. For our experiments we have chosen cobalt magnetic fluid 27-0001 from Strem Chemicals composed of 10 nm cobalt nanoparticles in kerosene coated with sodium dioctylsulfosuccinate and a monolayer of LP4 fatty acid condensation polymer. The average volume fraction of cobalt nanoparticles in this ferrofluid is p=8.2%. Cobalt behaves as an excellent metal in the long wavelength infrared range (LWIR), as evident by Fig. 3A: real part of refractive index, $n$, is much smaller than its imaginary part, $k$. Thus, real part of $\varepsilon$, $\mathrm{Re}\varepsilon = n^2 - k^2$, is negative, and its absolute value is much larger than its imaginary part, $\mathrm{Im}\varepsilon = -2nk$. Therefore, it is highly suitable for fabrication of hyperbolic metamaterials. Electromagnetic properties of these metamaterials may be understood based on the Maxwell-Garnett approximation via the dielectric permittivities $\varepsilon_m$ and $\varepsilon_d$ of cobalt and kerosene, respectively. Volume fraction of cobalt nanoparticles aligned into nanocolumns by external magnetic field depends on the field magnitude. Let us call this variable α(B). At very large magnetic fields all nanoparticles are aligned into nanocolumns, so that $\alpha(\infty) =$ p=8.2%. We will assume that at smaller fields the difference $\alpha(\infty)$-α(B) describes cobalt nanoparticles, which are not aligned and distributed homogeneously inside the ferrofluid. Dielectric polarizability of these nanoparticles may be included into $\varepsilon_d$ leading to slight increase in its value. Using this model, the diagonal components of the ferrofluid permittivity may be calculated using Maxwell-Garnett approximation as follows [23]:

$$\varepsilon_z = \alpha(B)\varepsilon_m + (1-\alpha(B))\varepsilon_d \tag{6}$$

$$\varepsilon_{xy} = \frac{2\alpha(B)\varepsilon_m\varepsilon_d + (1-\alpha(B))\varepsilon_d(\varepsilon_d+\varepsilon_m)}{(1-\alpha(B))(\varepsilon_d+\varepsilon_m) + 2\alpha(B)\varepsilon_d} \tag{7}$$

Calculated wavelength dependences of $\varepsilon_z$ and $\varepsilon_{xy}$ at $\alpha(\infty) =$ p=8.2% are plotted in Figs. 3CD. While $\varepsilon_{xy}$ stays positive and almost constant, $\varepsilon_z$ changes sign to negative around $\lambda$ = 3 μm. If the volume fraction of cobalt nanoparticles varies, this change of sign occurs at some critical value $\alpha_H$:

$$\alpha > \alpha_H = \frac{\varepsilon_d}{\varepsilon_d - \varepsilon_m} \tag{8}$$

The value of $\alpha_H$ as a function of wavelength is plotted in Fig.3B. This plot indicates that the original ferrofluid diluted with kerosene at a 1:10 ratio remains a hyperbolic medium above $\lambda$ = 5μm. More interestingly, such a diluted ferrofluid develops very pronounced phase separation into cobalt rich and cobalt poor phases if subjected to external magnetic field. Optical microscope images of the diluted ferrofluid before and after application of external magnetic field are shown in Figs. 1CD. The periodic pattern of self-assembled stripes visible in image D appears due to phase separation. The stripes are oriented along the direction of magnetic field. A movie taken during this experiment when magnetic field was turned on and off is available in the supporting on-line material. It clearly demonstrates the tunability of the medium by external magnetic field. The stripe periodicity $L$~2μm appears to be much smaller than the free space wavelength in the hyperbolic frequency range. Therefore, created self-assembled optical medium appears to be a photonic hyper-crystal. We should also note that the original undiluted ferrofluid exhibits similar phase separation in external magnetic field, even though on much smaller ~0.3 μm spatial scale (see Fig.1E).

Polarization dependencies of ferrofluid transmission as a function of magnetic field and nanoparticle concentration measured in a broad 0.5 μm - 16 μm wavelength range conclusively prove hyperbolic crystal character of ferrofluid anisotropy in the long wavelength IR range at large enough magnetic field. Fig. 3E shows polarization-dependent transmission spectra of 200 μm thick ferrofluid sample obtained using FTIR spectrometer. These data are consistent with hyperbolic character of $\varepsilon$ tensor of the ferrofluid in $B$ = 1000 G. Ferrofluid transmission is large for polarization direction perpendicular to magnetic field (perpendicular to cobalt nanoparticle chains) suggesting dielectric character of $\varepsilon$ in this direction. On the other hand, ferrofluid transmission falls to near zero for polarization direction along the chains, suggesting metallic character of $\varepsilon$ in this direction. However, these measurements are clearly affected by numerous ferrofluid absorption lines, whose behaviour will be addressed later. In order to be conclusive, LWIR data must be supported by polarization measurements in the visible range presented in Fig.4. Main features of these dependencies may indeed be understood based on the Maxwell-Garnett approximation (eqs.(6,7)). At small magnetic fields α(B)<<p and we obtain

$$\varepsilon_z = \alpha\varepsilon_m + (1-\alpha)\varepsilon_d \approx \varepsilon_d + i\alpha\varepsilon''_m \text{ and } \varepsilon_{xy} \approx \varepsilon_d \ , \quad (9)$$

where $\varepsilon''_m$ is the imaginary part of $\varepsilon_m$. This absorption-related anisotropy is primarily responsible for the polarization dependence of ferrofluid transmission. Results presented in Figs.4AB were obtained at $\lambda$ = 1.55 μm by measuring transmission of ferrofluid placed inside a $d$ =10 μm wide quartz cuvette (note that $\varepsilon_{quartz} \sim \varepsilon_d$ in the visible and near IR ranges). If ferrofluid inside the cuvette may be considered as a homogeneous medium, its transmission as a function of light polarization ϕ should behave as

$$T = T_{\min}\sin^2\phi + T_{\max}\cos^2\phi = T_{\max}\left(1 - \left(\frac{T_{\max}-T_{\min}}{T_{\max}}\right)\sin^2\varphi\right) \quad (10)$$

where $T_{\min}$ is transmission for polarization direction parallel to magnetic field and $T_{\max}$ is transmission for polarization direction perpendicular to magnetic field. Since electric field equals $E = E_0\exp\left(-i\frac{2\pi\sqrt{\varepsilon}}{\lambda}d\right)$ after passing through a ferrofluid layer of thickness $d$, the contrast between transmissions can be estimated using expressions for $\varepsilon_z$ and $\varepsilon_{xy}$ from eq.(9):

$$\ln\left(\frac{T_{\max}}{T_{\min}}\right) \approx \frac{2\pi d\alpha}{\varepsilon_d^{1/2}}\left(\frac{\varepsilon''_m}{\lambda}\right) \quad (11)$$

Equation (11) may be used to verify Maxwell-Garnett approximation and measure α(B). At $\lambda$ = 1.5 μm and large fields, polarization curves depart from the Malus law, showing much sharper minima (Fig. 4A), which will be discussed later. However, in the visible range, or in smaller fields, polarization curves indeed obey the Malus law exhibiting $\sin^2\phi$ dependence (see Fig. 4C). Moreover, low field measurements in Fig. 4D performed on the same sample at four different wavelengths across a very broad 0.5-10.6 μm wavelength range follow the Maxwell-Garnett result given by Eq. 11 (note that $\alpha$ = 0.0004 <<p in this case, since only a small fraction of particles are aligned into nanocolumns in a week magnetic field). These measurements give substantial support to our model.

Let us demonstrate that these measurements conclusively point towards hyperbolic character of ferrofluid anisotropy in the long wavelength IR range. As an example, let us calculate $\alpha(B)$ for the polarization curve in Fig. 4C measured at $B$ = 1630 G. While polarization contrast is quite large for this curve: $T_{max}/T_{min}$ ~ 10, it fits Malus law $\sin^2\phi$ dependence rather well. Based on Eq. 11, calculated $\alpha(B)$ equals 0.002 in this experiment performed at $\lambda$ = 488 nm. Therefore, at $\lambda$ = 14 μm where $\mathrm{Re}(\varepsilon_m)$ = -2100 the real parts of the dielectric permittivity tensor components are

$$\mathrm{Re}(\varepsilon_z) \approx \alpha\varepsilon'_m + \varepsilon_d \approx -2.2 \text{, while } \mathrm{Re}(\varepsilon_{xy}) \approx \varepsilon_d \approx +2.1 \qquad (12)$$

meaning that the magnetized ferrofluid does indeed exhibit hyperbolic metamaterial behaviour.

Polarization dependent transmission of phase separated ferrofluid samples shown in Fig. 1D further supports long range order of periodically aligned cobalt nanocolumns in a magnetized ferrofluid. It is clear from Fig.4A that polarization curves measured at large magnetic fields deviate strongly from the Malus law $\sin^2\phi$ dependence. Moreover, as shown in Fig.4B, polarization dependencies of the phase separated ferrofluid measured at different nanoparticle concentrations may be scaled to a universal "polarization notch" curve described by two parameters:

$$T \sim \frac{1 - a\sin^2\phi}{1 + b\cos^2\phi} \tag{13}$$

Such a “polarization notch” dependence arise naturally due to multiple scattering of light by periodically spaced nanoparticle rich and nanoparticle poor phases seen in Fig.1D. Let us consider light propagation through a single “Fabry-Perot resonator” formed by the phase boundaries. Transmission amplitude t through such an elementary resonator is given by a coherent sum of all the multiple scattering events in which the wave experiences multiple polarization-dependent reflections from the phase boundaries:

$$t = \sum_{m=0}^{\infty} t_m = t_0 \sum_{m=0}^{\infty} r^m e^{im\delta} = \frac{t}{1 - re^{i\delta}} \tag{14}$$

where δ is the phase accumulated by the wave upon double pass between the phase boundaries. Resulting transmission through a single resonator formed by the nanoparticle rich and nanoparticle poor phases is thus

$$T = tt^* = \frac{t^2}{1 + r^2 - 2r\cos\delta} \tag{15}$$

Since both t and r exhibit strong polarization dependence, the universal “polarization notch” curve given by eq.(13) finds natural explanation. Note that in order for this model to be valid, polarization state of light must be conserved during propagation, indicating a well-ordered ferrofluid state. Thus, our microscopy results in Fig.1C-E, and detailed examination of polarization-dependent transmission of the ferrofluid presented in Fig.4 strongly indicate both hyperbolic character of the medium in the long wavelength IR range, and its well-ordered periodicity. These data conclusively prove that the magnetized ferrofluid may indeed be considered as a self-assembled photonic hyper-crystal.

As summarized in Fig.5, fabricated photonic hyper-crystals exhibit all the typical behaviors associated with usual hyperbolic metamaterials. For example, Figs. 5AB demonstrate modulation of the metamaterial absorption lines due to hyperbolic order induced by the external magnetic field. Radiation lifetime engineering due to broadband divergence of the photonic density of states in hyperbolic metamaterials has been predicted in [16,24] and experimentally confirmed in [17]. Absorption lines of kerosene offer natural target for testing this effect in fabricated photonic hyper-crystals. Absorption spectra measured using FTIR spectrometer with and without external magnetic field are consistent with the decrease of the radiation lifetime of kerosene molecules in the hyperbolic state. Shorter lifetime leads to decrease in absorption line amplitude, which is detected in Fig.5B. Reduced reflection has been reported recently as another experimental signature of hyperbolic metamaterials [25]. Due to broadband divergence of photonic density of states, roughened surface of a hyperbolic metamaterial scatters light preferentially inside the medium, resulting in reduced reflectance. In case of photonic hyper-crystals, periodic modulation of the hyper-crystal surface plays the same role, facilitating coupling of external light into the large k vector modes of photonic hyper-crystal. Figs. 5CD demonstrate such magnetic field induced reduced reflectivity of the ferrofluid.

Finally, we would like to illustrate photonic hyper-crystal potential in chemical and biological sensing, which is made possible by spatially selective field enhancement effects described above. This potential is revealed by detailed measurements of magnetic field induced transmission of photonic hyper-crystals in the broad IR spectral range presented in Figs. 6AB. FTIR spectral measurements are broadly accepted as a powerful "chemical fingerprinting" tool in chemical and biosensing. Therefore, broadly available magnetic field-tunable photonic hyper-crystals operating in the IR range open up new valuable opportunities in chemical analysis. Our experimental data presented in Fig. 6 clearly illustrate this point. The FTIR transmission spectrum of the diluted (p=0.8%) ferrofluid in Fig.6A exhibits clear set of kerosene absorption lines, which is consistent with other published data (see for example ref.[26]). On the other hand, magnetic field induced transmission spectrum of the p=8.2% ferrofluid shown in Fig.6B contains a very pronounced absorption line at $\lambda$~12 μm (~840 $cm^{-1}$), which cannot be attributed to kerosene. Quite naturally this absorption line may be attributed to fatty acids, since cobalt nanoparticles are coated with a monolayer of surfactant composed of various fatty acids, such as lactic, oleic etc. acids as shown in Fig.6D. Detailed comparison of Fig.6B with the 10-14 μm portion of lactic acid FTIR absorption spectrum shown in Fig.6C indeed indicates a close match. The fatty acid line appears to be about as strong as $\lambda$~14 μm (~695 $cm^{-1}$) line of kerosene, even though the oscillator strength of these molecular lines is about the same, while the amount of kerosene in the sample is ~ two orders of magnitude larger (a monolayer coating of fatty acids on a 10 nm cobalt nanoparticle occupies no more than 1% of ferrofluid volume). This paradoxical situation clearly indicates local field enhancement effects. Another strong evidence of field enhancement is provided by measurements of extinction coefficient of the ferrofluid presented in Fig.6E. Ferrofluid subjected to magnetic field exhibits pronounced resonances around the fatty acid absorption line at $\lambda$~12 μm and the kerosene absorption line at $\lambda$~14 μm. These resonances provide clear evidence of field enhancement by cobalt nanoparticle chains. We expect that further optimization of photonic hyper-crystals geometry will lead to much stronger sensitivity of their optical properties to chemical and biological inclusions, indicating a very strong potential of photonic hyper-crystals in biological and chemical sensing.

**Acknowledgement**

This work was supported in part by NSF grant DMR-1104676, NSF Center for Photonic and Multiscale Nanomaterials, ARO MURI and Gordon and Berry Moore Foundation.

## Figure Captions

**Fig. 1**. (A) Typical geometry of metal nanowire-based hyperbolic metamaterials. (B) An example of a photonic hyper-crystal: since photon wave vector in hyperbolic metamaterials is not diffraction-limited, periodic modulation of hyperbolic metamaterial properties on a scale $L<<\lambda$ would lead to Bragg scattering and formation of band structure no matter how small L is. (C-E) Microscopic images of cobalt nanoparticle-based ferrofluid reveal subwavelength modulation of its spatial properties: frames (C) and (D) show microscopic images of the diluted cobalt nanoparticle-based ferrofluid before and after application of external magnetic field. The pattern of self-assembled stripes visible in image D is due to phase separation of the ferrofluid into cobalt reach and cobalt poor phases. The stripes are oriented along the direction of magnetic field. A movie taken during this experiment when the external magnetic field was turned on and off is available in the supporting on-line material. Frame (E) demonstrates that the original undiluted ferrofluid exhibits similar phase separation in external magnetic field, even though on much smaller scale.

**Fig. 2.** Comparison of the effective medium dispersion of a phonic hyper-crystal (A) to the exact solution (B). The hyper-crystal unit cell is formed by 250 nm of $In_{0.53}Ga_{0.47}As$:$Al_{0.48}In_{0.52}As$ semiconductor hyperbolic metamaterial [18] with 5μm plasma wavelength, followed by 250 nm dielectric layer of $Al_{0.48}In_{0.52}As$. $k_0$ is the free-space wavenumber.

**Fig. 3**. (A) Optical properties of cobalt as tabulated in ref.(18): real n and imaginary k parts of the cobalt refractive index are plotted in the long wavelength IR range. (B) Critical volume fraction of cobalt nanoparticles corresponding to ferrofluid transition to hyperbolic metamaterial phase plotted as a function of free space light wavelength.

(C,D) Wavelength dependencies of $\varepsilon_z$ and $\varepsilon_{xy}$ at $\alpha(\infty)$=8.2%. While $\varepsilon_{xy}$ stays positive and almost constant, $\varepsilon_z$ changes sign to negative around $\lambda$=3μm. (E) Polarization-dependent transmission spectra of 200 μm thick ferrofluid sample measured using FTIR spectrometer are consistent with hyperbolic character of ε tensor.

**Fig. 4**. (A,B) Experimentally measured transmission of the cobalt based ferrofluid at $\lambda$=1.55 μm as a function of cobalt concentration, external magnetic field and polarization angle. As shown in panel (B), in the absence of phase separation polarization curves exhibit the Malus law $\sin^2\phi$ dependencies. On the other hand, in the phase separated state polarization curves exhibit universal "polarization notch" behaviour indicating long range order of periodically aligned cobalt nanocolumns. (C) Experimentally measured transmission of the ferrofluid at $\alpha(\infty)$ = 8.2%, $\lambda$ = 488 nm and B = 1630 G. (D) Polarization contrast measurements (data points) in the 0.5-10.6 μm range provide experimental validation of Maxwell-Garnett approximation (red line) at low nanoparticle concentrations.

**Fig. 5.** (A) FTIR transmission spectrum of the diluted ($\alpha(\infty)$=0.8%) ferrofluid in zero magnetic field reveals strong kerosene absorption lines. (B) Measured differential FTIR transmission spectrum T(H)-T(0) indicate decrease in absorption line amplitude, which is consistent with the decrease of the kerosene radiation lifetime in the hyperbolic metamaterial state. (C) FTIR reflection spectrum of the ferrofluid in zero magnetic field. (D) Measured differential FTIR reflection spectrum R(H)-R(0) indicates reduced reflectivity in the hyperbolic state.

**Fig. 6**. (A) FTIR transmission spectrum of diluted ($\alpha(\infty)$=0.8%) ferrofluid exhibits clear set of kerosene absorption lines. (B) Transmission spectra of the $\alpha(\infty)$=8.2%

ferrofluid measured with and without application of external magnetic field. Magnetic field induced transmission spectrum contains a very pronounced absorption line at $\lambda$~12 $\mu$m (~840 $cm^{-1}$), which can be attributed to lactic acid. Kerosene absorption lines are marked with yellow boxes, while the fatty acid line at 840 $cm^{-1}$ is marked with a green box. (C) The 10-14 $\mu$m portion of lactic acid FTIR absorption spectrum. (D) Schematic view of cobalt nanoparticle coated with a monolayer of fatty acids, such as lactic and oleic acid. (E) Extinction coefficient of ferrofluid subjected to external magnetic field exhibits pronounced resonances around the fatty acid absorption line at $\lambda$~12 $\mu$m and the kerosene absorption line at $\lambda$~14 $\mu$m. These resonances provide clear evidence of field enhancement by cobalt nanoparticle chains.

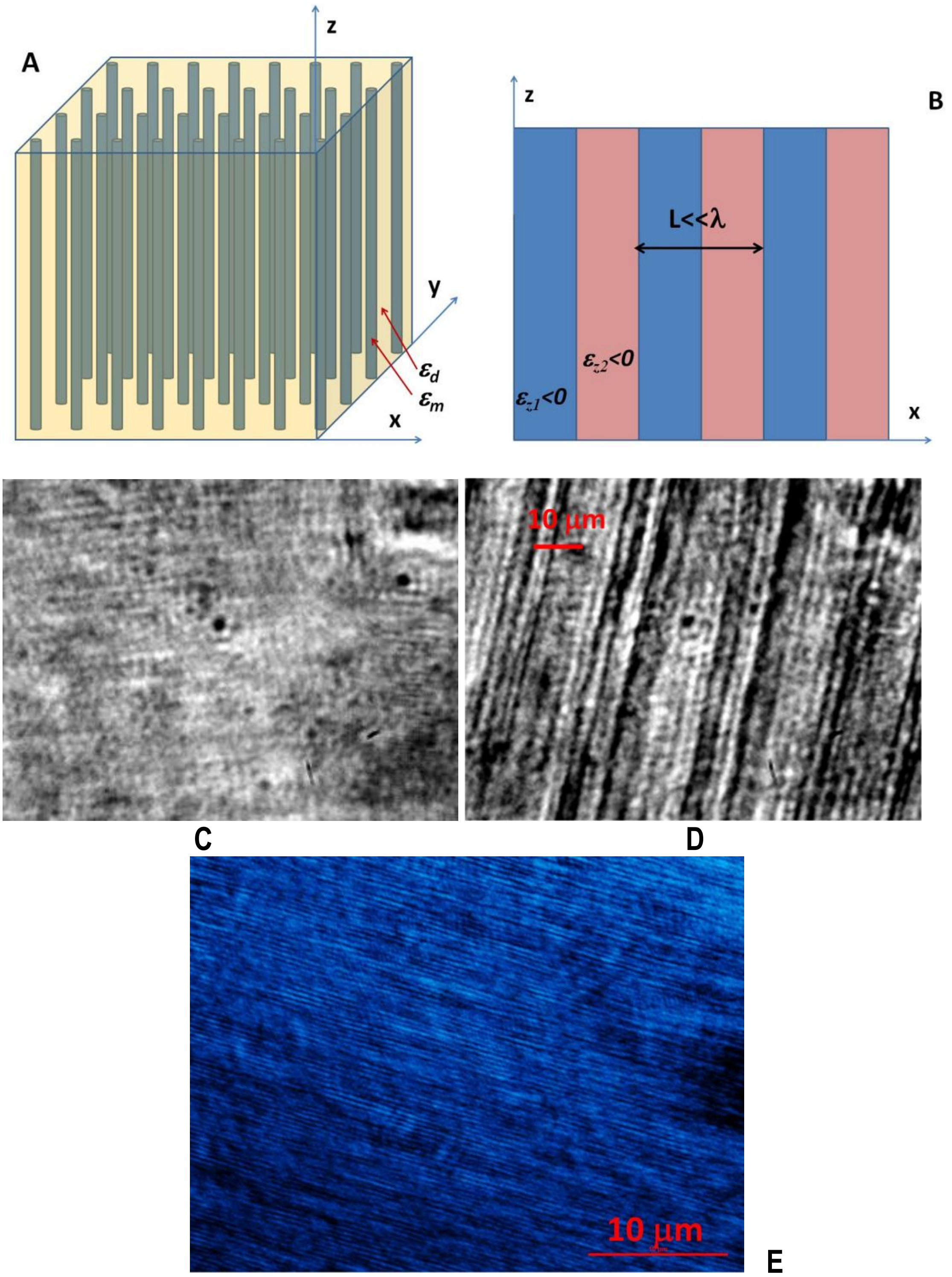

**Fig. 1**

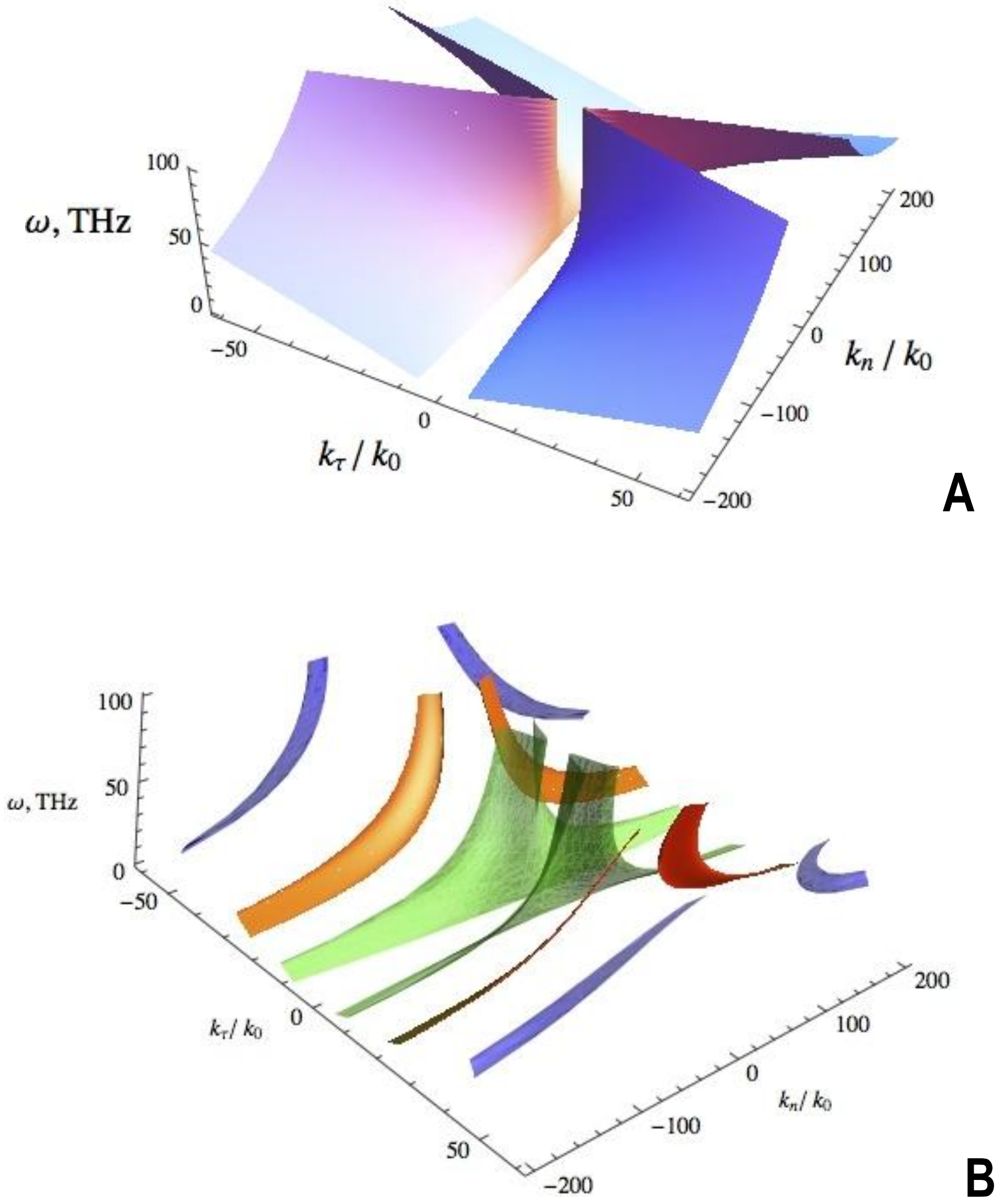

**Fig. 2**

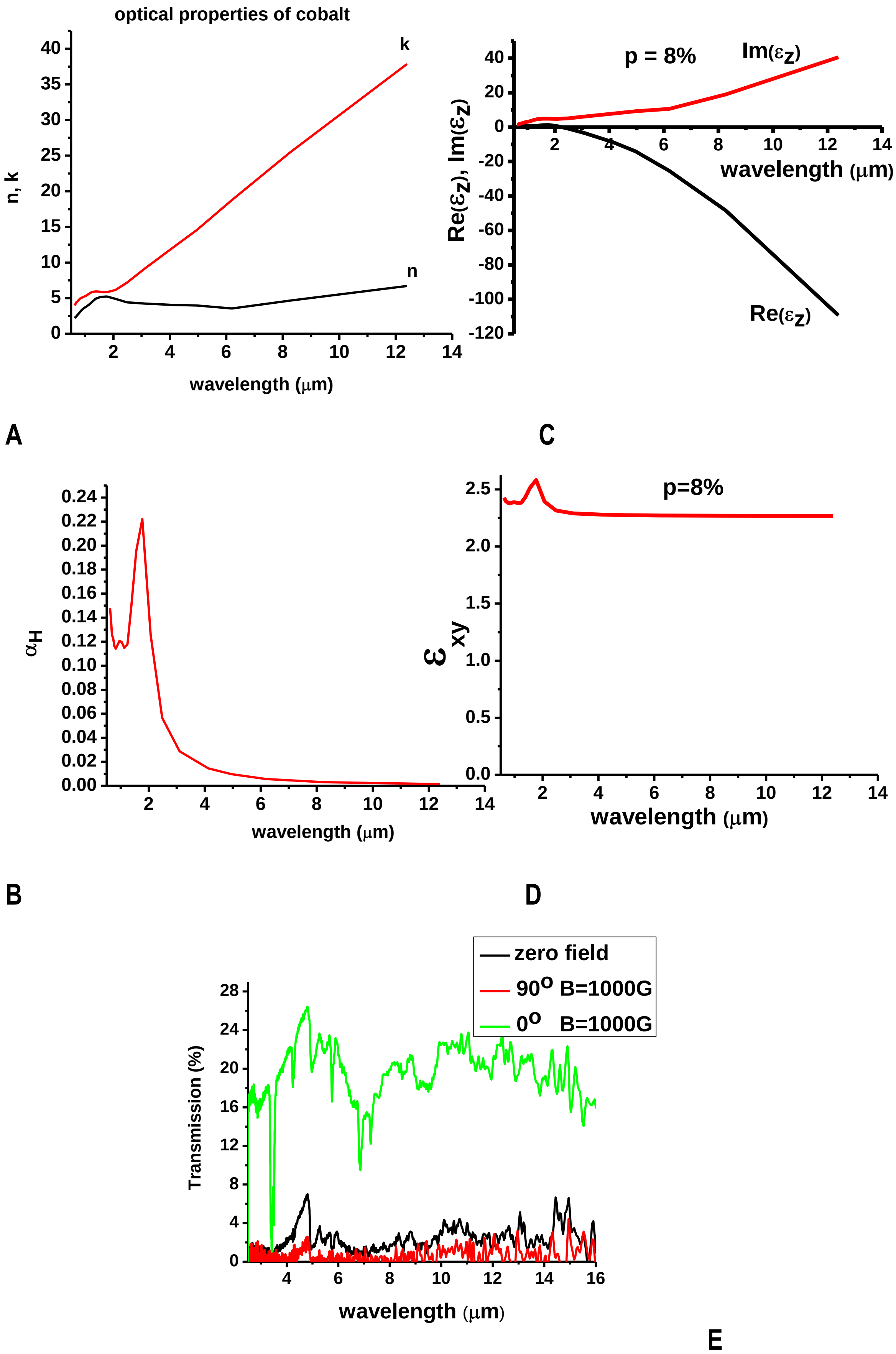

Fig. 3

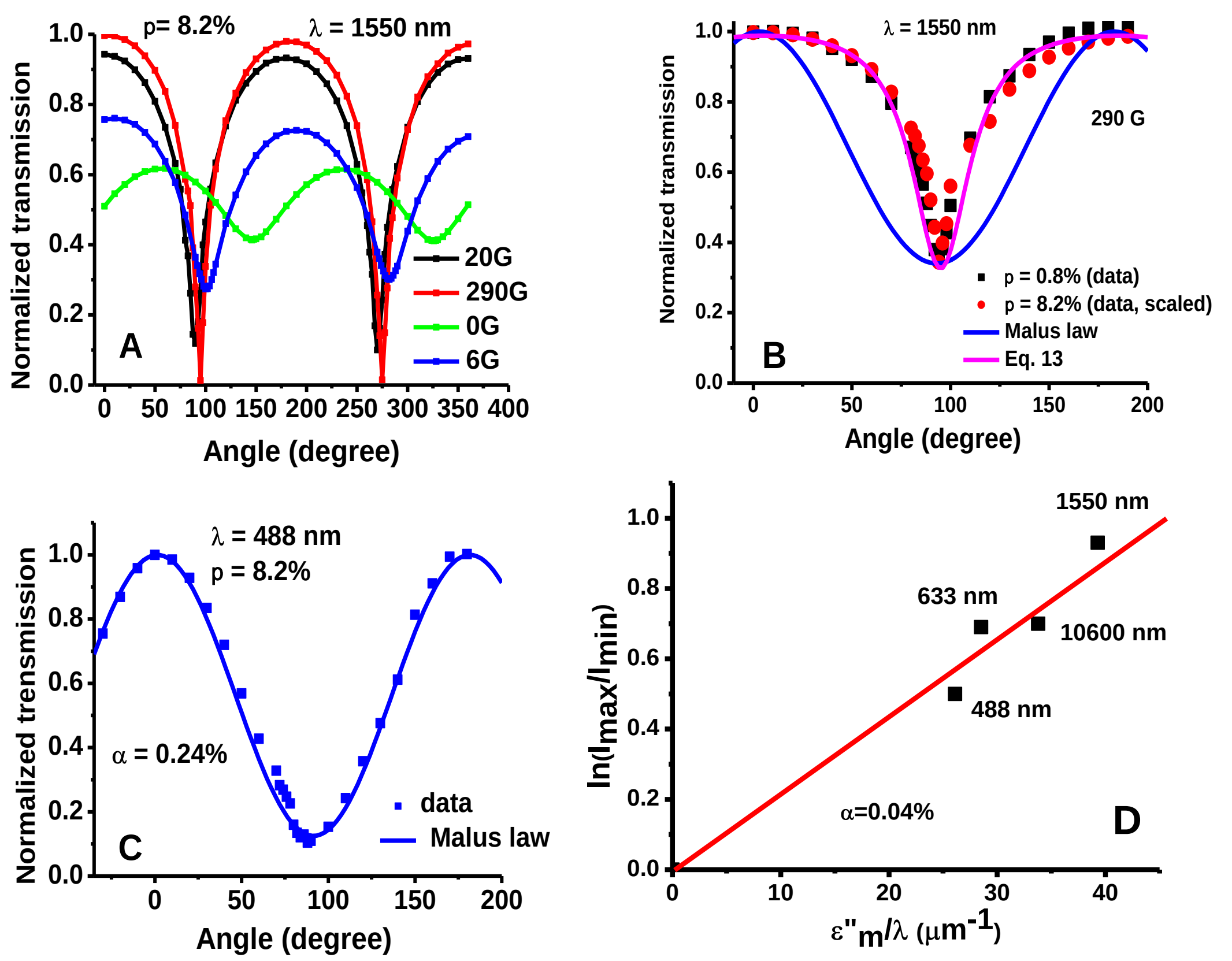

Fig. 4

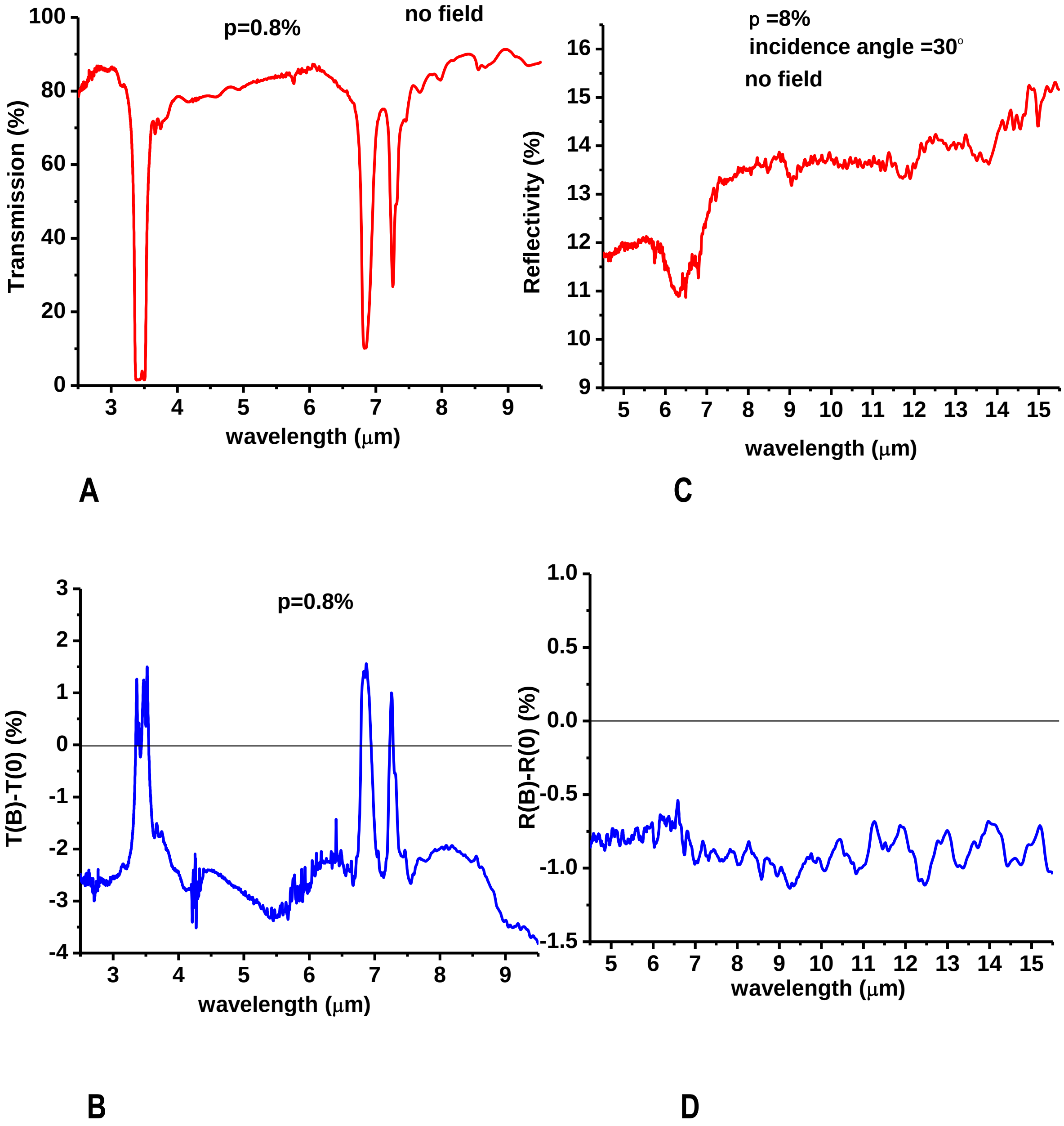

Fig. 5

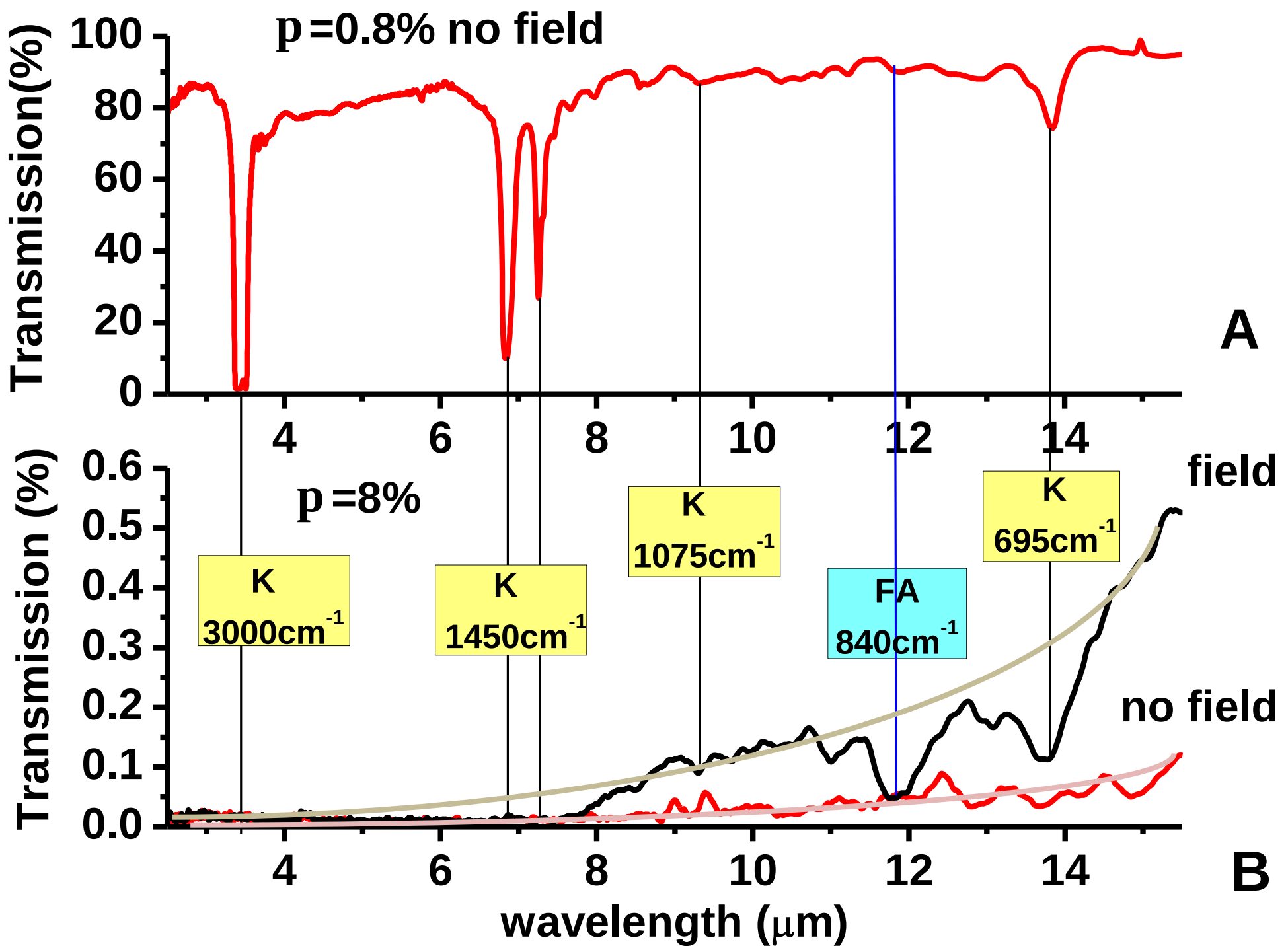

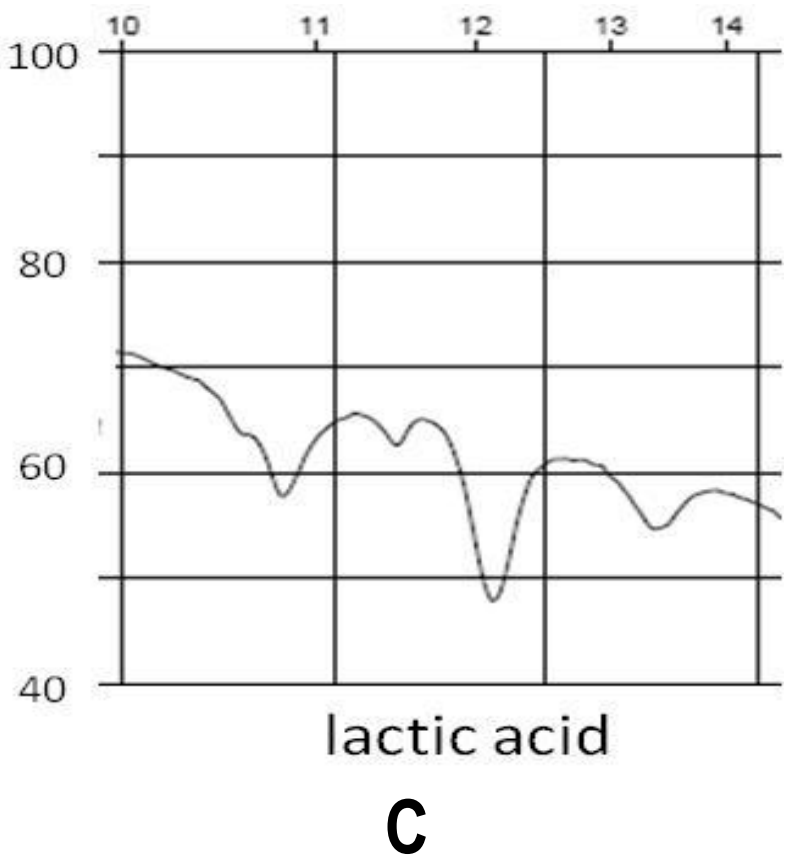

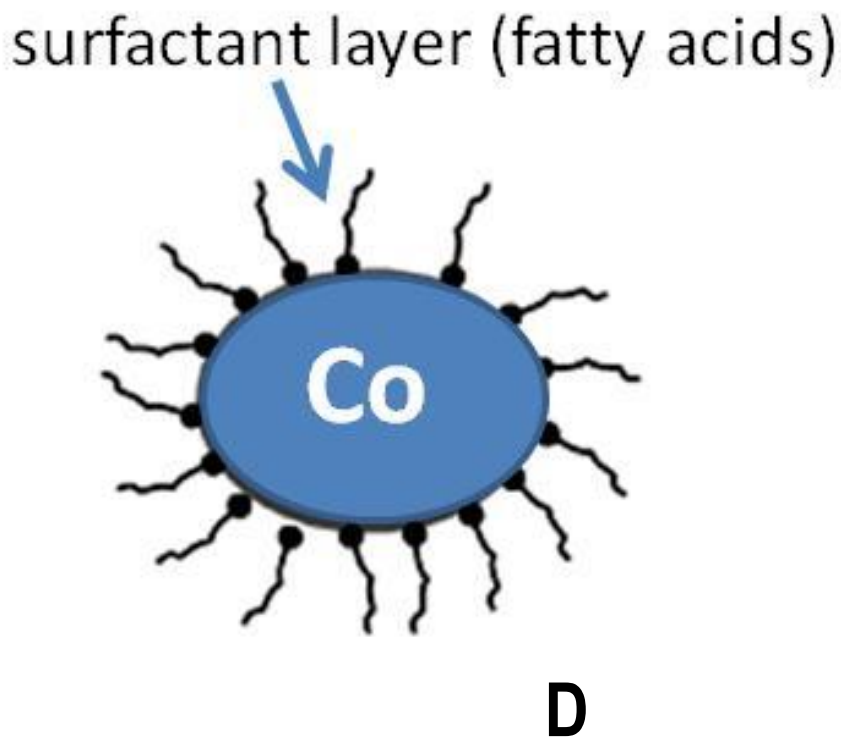
surfactant layer (fatty acids)
Co
D

Fig. 6

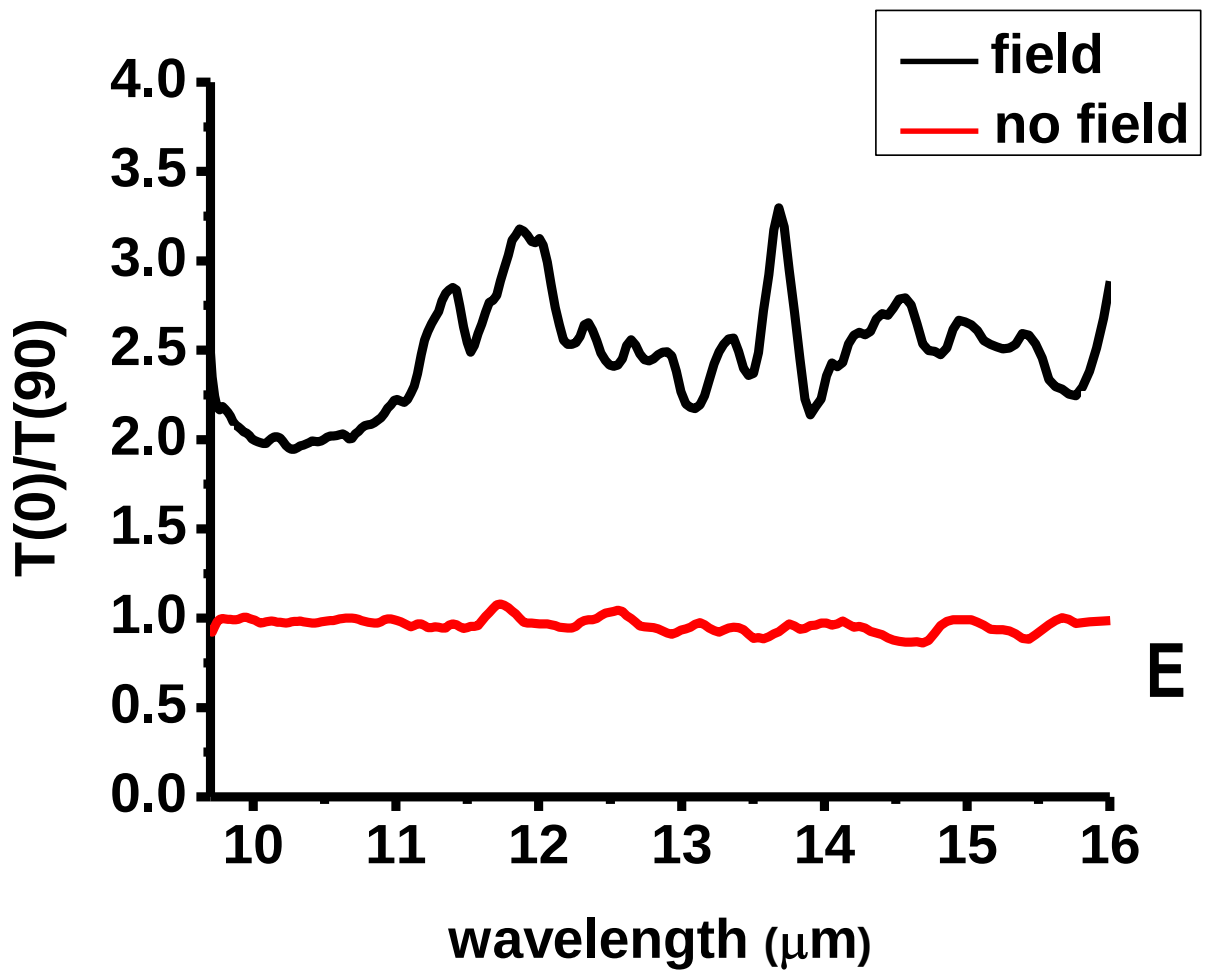